\documentclass[12pt]{iopart}
\begin{document}

\title[Lattice distortion and orbital ordering to resonant x-ray
scattering in manganites]{Relative contributions
of lattice distortion and orbital ordering to resonant x-ray scattering
in manganites} 

\author{M Nakamura$^{1}$, M Izumi$^{2}$, N Ogawa$^{1}$,
H Ohsumi$^{3}$, Y Wakabayashi$^{4}$ and K Miyano$^{2}$}

\address{$^{1}$ Department of Applied Physics, University of Tokyo, Tokyo
113-8656, Japan }

\address{$^{2}$ Research Center for Advanced Science and Technology,
University of Tokyo, Tokyo 153-8904, Japan}

\address{$^{3}$ Japan Synchrotron Radiation Research Institute (JASRI),
Mikazuki, Hyogo 679-5198, Japan}

\address{$^{4}$ Photon Factory, Institute of Materials Structure Science, High
Energy Accelerator Research Organization, Tsukuba 305-0801, Japan}

\ead{masao-n@myn.rcast.u-tokyo.ac.jp}

\begin{abstract}
We investigated the origin of the energy splitting observed in the
resonant x-ray scattering (RXS) in manganites. Using thin film samples
with controlled lattice parameters and orbital states at a fixed orbital
filling, we estimated that the contribution of the interatomic Coulomb
 interaction relative to the Jahn-Teller mechanism is insignificant and at
 most 0.27. This indicates that RXS probes mainly Jahn-Teller distortion
 in manganites. 
\end{abstract}


\submitto{\JPCM}


\nosections
Hole-doped manganites with perovskite structure,
$R_{1-x}A_{x}$MnO$_{3}$ ($R$ is a trivalent rare-earth ion and $A$ is
a divalent alkaline-earth ion) show a variety of phases with doping
concentration $x$. In these compounds, double-exchange interaction has
been considered most important, which closely couples
spin with charge degree of  freedom. However, it has been recently
recognized that double-exchange alone is not enough to account for the
peculiar properties of manganites. Theoretical and experimental studies
have shown the importance of the orbital degree of freedom to understand
the rich phase diagrams~\cite{tokura,maezono}.    

The direct information of the orbital ordering has been lacking due to the
dearth of proper experimental means until Murakami \textit{et al.} 
proposed the use of resonant x-ray scattering (RXS)\cite{murakami1,murakami2}.
Anisotropic charge distribution in an atom
results in the anisotropy in the x-ray susceptibility near the
absorption edge. This gives rise to x-ray dichroism and scattering
intensity at structurally forbidden Bragg reflections for polarized
x-ray~\cite{templeton,dmitrienko}. They exploited this technique and
succeeded in detecting the resonant scattering from superlattice
reflections, which has been interpreted as a signature of the
antiferro-type orbital ordering. The resonance at main edge corresponds
to the dipolar transition from $1s$ to unoccupied $4p$ level (Mn
$K$-edge), and the split of the $4p$ levels is the origin of the anisotropy.
It has been argued that the RXS reflects the $3d$ orbital ordering
because the anisotropic $3d$ charge distribution causes the $4p$
splitting via the Coulomb repulsion~\cite{ishihara}. However, this
interpretation has been later challenged; the $4p$ splitting is the result
of Jahn-Teller distortion rather than the $3d$ orbital
polarization~[8 - 12].
There has been no definitive answer to these conflicting views.

In order to distinguish the two mechanisms, we have performed RXS
measurements of samples with controlled lattice constant and orbital
ordering at a fixed orbital filling (i.e., the average number of
electrons occupying the $e_g$ orbital). An epitaxially grown film is suitable
for this purpose because the epitaxial strain makes possible to
artificially modify the lattice constant. Thin films in this paper
were all deposited on SrTiO$_{3}$ (001) substrate and their
hole concentration was fixed to 0.4, i.e., 0.6 in the orbital
filling. The composition of the films was $R_{0.6}$Sr$_{0.4}$MnO$_{3}$, where
$R$ was La, La$_{0.6}$Pr$_{0.4}$, La$_{0.2}$Pr$_{0.8}$, Pr,
Pr$_{0.6}$Nd$_{0.4}$, or Nd in the decreasing order of average ionic
size with correspondingly shorter out-of-plane lattice constant
($c$-axis) ({\it vide infra}). The in-plane lattice constant ($a$-axis)
was fixed to that of the substrate. Bulk crystals of identical composition
are all ferromagnetic metal (FM) in the ground
state~\cite{tomioka}. Because the modulation of orbital states alters
the magnetic, transport, and optical properties of thin film
manganites~[14 - 16], any deviation 
from FM in our film signals the change in the orbital state. 
The lattice constant of SrTiO$_{3}$ is 3.905 \AA, which is larger than
those of these compounds in the bulk. Our epitaxial films were thus
laterally stretched causing a split in the doubly degenerate $e_{g}$
orbitals into $x^{2}-y^{2}$ at lower level and $3z^{2}-r^{2}$ at higher
level. Therefore, one probable orbital state is the $x^{2}-y^{2}$-type
orbital order accompanied by the layer-type (A-type) antiferromagnetic
(A-AF) state, typified by bulk crystals of
Nd$_{1-x}$Sr$_{x}$MnO$_{3}$~\cite{kajimoto,kuwahara}. 
This is because the orbital ordering promotes in-plane transport,
which brings about in-plane ferromagnetic coupling by double-exchange
interaction and out-of-plane antiferromagnetic coupling by
superexchange interaction. As will be shown later, the
magnetic and magnetotransport measurements indicate that we have indeed
films ranging from FM to A-AF.

RXS measurements were performed at beam-line 4C and 9C at Photon
Factory, KEK, Tsukuba, employing newly-developed interference
technique by Kiyama~\textit{et al.}~\cite{kiyama}. We observed energy
dependence of interference term from (102) reflection of
La$_{0.6}$Sr$_{0.4}$MnO$_{3}$ (LSMO), Pr$_{0.6}$Sr$_{0.4}$MnO$_{3}$
(PSMO) and Nd$_{0.6}$Sr$_{0.4}$MnO$_{3}$ (NSMO) films at various
azimuthal angles with the photon energy near the Mn $K$-absorption edge.
Our results show that the Jahn-Teller mechanism has dominant influence
on Mn $4p$ splitting, which is consistent with other reports for
manganite thin films~\cite{kiyama,ohsumi,song}. 

Thin film samples were grown using pulsed laser deposition
technique~\cite{izumi}. The thickness of the films was about 50~nm.
X-ray diffraction measurements for $(00l)$ and $(114)$
reflection peaks confirmed that all films grew
epitaxially with a (001) orientation and that the out-of-plane lattice
constant systematically decreased as the $A$-site ion was replaced by a
smaller ion as depicted in the abscissa in \fref{fig1}(b).
\Fref{fig1}(a) shows the temperature dependence of the resistivity
measured within the $ab$-plane using a common four probe method and (b)
shows the magnetization at 5~K under the magnetic field of 500~Oe after
field cooling. All films had metallic ground state and the insulator to metal
(I-M) transition temperature ($T_{\mathrm{IM}}$) was continuously
shifted to lower temperatures from the LSMO to the NSMO film.  
However, the magnetic property exhibited a dramatic variation with the
lattice constant ratio $c/a$. The LSMO film had ferromagnetic transition
at 325~K and the magnetization was about 3.2~$\mathrm{\mu_{B}/Mn}$ in
the ground state. As $c/a$ is reduced, the magnetization decreased; for 
the NSMO film, the magnetization was only about
0.26~$\mathrm{\mu_{B}/Mn}$ and didn't saturate even at 50 kOe. The small
spontaneous magnetization in spite of in-plane metallic conductivity
signals the A-AF ordering. However one might still argue that the small
magnetization and the metallic conduction are the result of the phase
separation rather than the A-AF ordering: small fraction of
ferromagnetic metallic region is embedded in antiferromagnetic
insulator. In order to differentiate the two possibilities we measured
the transport across a NSMO film. 

A NSMO film was deposited on 5\% La-doped SrTiO$_{3}$ (La-STO), which is
a good conductor. By x-ray diffraction and $M$-$T$ and $M$-$H$ curves, we
confirmed that it had the same lattice constant and magnetic properties as
those of the NSMO film grown on STO. The film was etched to form two
$100\times 100~\mu$m square pads separated by a 10~$\mu$m gap using
photolithography technique (see the inset of \fref{fig2}). Because
the substrate is highly conductive, the resistivity was estimated as
that coming from two 70~nm thin film resistors in series. We show in
\fref{fig2} the result of temperature dependence of out-of-plane
resistivity of the NSMO/La-STO film measured under the magnetic field of
0 and 50~kOe respectively applied parallel 
to the $ab$-plane. The large out-of-plane resistivity could be partly
due to the insulating layer formation at the interface, which
contributes to the overall offset of the resistivity. However, the
interfacial layer is inactive electronically in general and does not
respond to the magnetic field in particular~\cite{bibes}. Therefore, the 
appearance of the out-of-plane magnetoresistance around 200~K
(\fref{fig1}(b)) concomitant with the in-plane I-M transition in
\fref{fig1}(a), which persists down to low temperatures, signifies
the A-AF-type order inside the film. The negative magnetoresistance is
brought about by the canting of the out-of-plane antiparallel spin
order. This behaviour is consistent with the observations in the  bulk
crystals~\cite{kuwahara}. For a phase separated film, the magnetoresistance
should peak around the Curie temperature.  

While the magnetotransport measurement showed the evidence of
ferro-type orbital ordering in NSMO, the LSMO film should have more
isotropic orbital distribution judged from the 3D FM behaviour. 
Thus we have samples with varing degree of orbital polarization and a
range of $c/a$ values, which allow us to differentiate the two
mechanisms. 

In the previous RXS measurements used for antiferro-type
orbital ordering  in manganites, the incident beam is $\sigma$-polarized
and only the $\pi'$-polarized scattering beam is detected at a superlattice
position. In the present case, however, the diffraction spots from the
ferro-type orbital ordering and/or the Jahn-Teller distortion coincide
with the Bragg peaks. Thus the quite large $\sigma\to\sigma'$ Bragg
scattering masks the $\sigma\to\pi'$ scattering due to the anisotropic
scattering factor of Mn atoms, the information we are after.
 To circumvent the difficulty, an interference technique was
devised~\cite{kiyama,ohsumi}.  

The new technique is a method to extract an interference term between
$\sigma\to\sigma'$ scattering and $\sigma\to\pi'$ scattering.
Consider a geometry in which the analyzer
crystal is deliberately rotated by $\Delta_{\varphi}$ from the angle at
which only the $\pi'$-polarized beam can pass. There is a small
projection of $\sigma\to\sigma'$ scattering onto the analyzer which
can interfere with the projection of $\sigma\to\pi'$ scattering.
Note that Bragg scattering from incident $\pi$-polarized x-ray
$(\pi\to\pi'~\mathrm{scattering})$ is negligible because
the incident x-ray is almost completely $\sigma$-polarized.
The interferrence effect is analysed as follows. The atomic scattering
factor of Mn atom is given by 
\begin{equation}
\hat{f}=\left(
  \begin{array}{@{\,}ccc@{\,}}
   f_{a} & 0 & 0\\
   0 & f_{a} & 0\\
   0 & 0 & f_{c} 
  \end{array}\right),
\end{equation}
where a uniaxial symmetry is assumed and the $z$-axis is taken along the 
$c$-axis of the film. Both $f_{a}$ and $f_{c}$ contain contribution
from resonant and non-resonant processes but the difference $\Delta
f=f_{c}-f_{a}$ contains the resonant scattering only. Note that $|\Delta 
f|\ll |f_{a}|,|f_{c}|$. The scattering amplitude $F_{\sigma '\sigma}$ of 
$\sigma \to \sigma'$ scattering is given by 
\begin{equation}
 F_{\sigma'\sigma}=F_{hkl}+(f_{c}-f_{a})\sin^{2}\alpha
\sin^{2}\Psi + f_{a},
\end{equation}
and that of $\sigma\to\pi'$ scattering, $F_{\pi'\sigma}$, is given by
\begin{equation}
 F_{\pi'\sigma}=(f_{c}-f_{a})\{\sin^{2}\alpha\sin\theta\sin\Psi\cos\Psi-\sin\alpha\cos\alpha\cos\theta\sin\Psi\},
\end{equation}
where $\alpha$ is the angle between the $c$-axis of the film and the
scattering vector, $\theta$ the scattering angle, $\Psi$ the azimuthal
angle~\cite{kiyama}. The structure factor due to all atoms other than
Mn, $F_{hkl}$, was calculated using the known atomic scattering factors
and assuming cubic symmetry. Note that Mn atom was put at the center of
a unit cell when calculating the structure factor, so the phase factor
of Mn atom equal unity.
When the analyzer is set at an angle $\varphi_{A}$ from the
$\sigma\to\sigma'$ channel, the scattering intensity is given by
\begin{eqnarray}
 I(\varphi_{A})\propto & |F_{\sigma'\sigma}\cos\varphi_{A}-F_{\pi'\sigma}\sin\varphi_{A}|^{2}\nonumber\\
& +|F_{\sigma'\sigma}\sin\varphi_{A}+F_{\pi'\sigma}\cos\varphi_{A}|^{2}\cos^{2}2\theta_{A}.
\end{eqnarray}
By taking the difference between the intensities at two symmetrically
located $\varphi_{A}$'s, we obtain
\begin{equation}
 I(\textstyle\frac{\pi}{2}+\Delta_{\varphi})-I(\textstyle\frac{\pi}{2}-\Delta_{\varphi})\propto 2Re[F_{\sigma'\sigma}F_{\pi'\sigma}^{*}]\sin^{2}2\theta_{A}\sin 2\Delta_{\varphi}.
\end{equation}
In our measurement, $2\theta_{A}\sim90^{\circ}$ and $\Delta_{\varphi}$
was set at $20^{\circ}$. The interference term can be approximated to
the first order of $\Delta f$ as,
\begin{equation}
 Re[F_{\sigma'\sigma}F_{\pi'\sigma}^{*}]\sim Re[(F_{hkl}+f_{a})\cdot\Delta f^{*}]\label{eq:6}.
\end{equation}

\Fref{fig3} shows the energy dependence of the normalized intensity
of the interference term at azimuthal angle $\Psi=90^{\circ}$ at room
temperature for NSMO film. The error bar indicates one standard
deviation. The data are fitted using equation~(\ref{eq:6}), in which the
energy dependence of $\Delta f$ dominates over that of $F_{hkl}+f_{a}$.
$\Delta f$ is expressed as $\Delta f\sim f(E+\Delta E)-f(E)$
where, $\Delta E=E(4p_{z})-E(4p_{x})$ is the fitting parameter.
Since $f(E+\Delta E)-f(E)\sim (df/dE) \Delta E$, the shape of the
interference spectrum is predetermined by the derivative $df/dE$
and the fitting is sensitive 
only to the depth of the dip. Therefore the error bar of the data
directly translates into the error bar in the fitting parameter (shown
in \fref{fig4} below).
In the fitting procedure, the atomic scattering factor $f$ of Mn atom
was deduced from the Kramers-Kronig transformation of the absorption
spectrum and approximated to be equal to $f_{a}$. $f$ was numerically
differentiated to arrive at $df/dE$.
We determined $\Delta E=1.45\pm0.1$~eV for the NSMO film. 
The azimuthal dependence of the interference intensity is shown in the
inset of \fref{fig4} with a satisfactory fit again with $\Delta
E=1.45$~eV.  The relation of the energy levels is $E(4p_{z})<E(4p_{x})$ in
Coulomb mechanism and $E(4p_{z})>E(4p_{x})$ in Jahn-Teller mechanism.
The result $\Delta E>0$ implies that the Jahn-Teller effect is stronger.
The magnitude of the energy split for LSMO and PSMO at room temperature
were also estimated in a similar way.
\Fref{fig4} shows the relation between $c/a$ and the energy gap.
$\Delta E$ changes linearly with $c/a$, although
the orbital states of the three samples are expected to be considerably
different as will be discussed below.

We also performed measurement at low temperatures with a closed-cycle He
refrigerator for the NSMO film. 
The NSMO film has the transition from paramagnetic to A-AF ordering at 220~K
($T_{N}$), which promotes the shift of the orbital state to 
the $x^{2}-y^{2}$ orbital, through the spin-orbit interaction.
However, as is shown in \fref{fig3}, no noticeable change in $\Delta 
E$ was found above and below (at 150~K) $T_{N}$, indicating that the
effect of Coulomb mechanism is very small.

Detailed first-principle band structure calculations are available to
estimate the occupancy of the $e_{g}$ orbitals ($n_{x^{2}-y^{2}}$,
$n_{3z^{2}-r^{2}}$) as a function of the $c/a$ ratio and the spin
state~\cite{fang,fang2}. Application of the results of the calculations
to the samples in the current experiment leads to the following
estimates for the orbital polarization defined as
$\Delta n=n_{x^{2}-y^{2}}-n_{3z^{2}-r^{2}}$ for a fixed total occupancy
of 0.6; 
(i) $\Delta n=0.064$ for $c/a=0.969$ and A-AF phase
(the case of the NSMO at 150~K) and
(ii) $\Delta n=0.014$ for $c/a=0.981$ and ferromagnetic phase
(LSMO at room temperature).
Let us assume that the observed splitting $\Delta E$ is the sum of 
the contribution from the Jahn-Teller mechanism ($\Delta
E_{\mathrm{JT}}$: proportional to the magnitude of the
lattice distortion, or $1-c/a$) and that from the Coulomb mechanism
($\Delta E_{\mathrm{C}}$: proportional to $\Delta n$); 
$\Delta E=\Delta E_{\mathrm{JT}}-\Delta E_{\mathrm{C}}$.
Two data points in \fref{fig4} for LSMO and NSMO at 150~K then lead
to two simultaneous linear equations with two unknowns, $\Delta
E_{\mathrm{JT}}$ and $\Delta E_{\mathrm{C}}$. We obtain $\Delta
E_{\mathrm{C}}/\Delta E_{\mathrm{JT}}=0+0.27$ allowed within the error
bars of the two data points. The negative value is
of course non-physical. The worst possible value +0.27 corresponds to
the case in which $\Delta E$ values at two opposite ends of the error
bars (indicated by arrows in \fref{fig4}) are used; a highly unlikely
situation. Thus the above estimates must be quite
conservative. The fairly good fit of the line that passes through
$\Delta E=0$ at $c/a=1$ irrespective of the orbital states 
attests the smallness of the Coulomb contribution~\cite{note}.

In conclusion, we investigated the mechanism of RXS 
with use of $R_{0.6}$Sr$_{0.4}$MnO$_{3}$ thin films epitaxially grown on
SrTiO$_{3}$ substrates. The samples cover a wide range of orbital states,
to which the results of RXS are quite insensitive. The
upper limit of 0.27 is estimated for the relative 
contributions of the Coulomb and the Jahn-Teller mechanisms.

\ack
The authors thank Y. Murakami, H. Nakao, and M. Kubota for assistance in the
course of the RXS measurements and for fruitful discussions.
We are also grateful to Z. Fang for providing us with the data for
calculation. The work was supported in part by a Grand-in-Aid for COE
Research from the MEXT of Japan.

\Figures
\begin{figure}
 \caption{\label{fig1}Temperature dependence of the resistivity of
 $R_{0.6}$Sr$_{0.4}$MnO$_{3}$ ($R$=La, La$_{0.6}$Pr$_{0.4}$,
 La$_{0.2}$Pr$_{0.8}$, Pr, Pr$_{0.6}$Nd$_{0.4}$, and Nd)
films on SrTiO$_{3}$ substrate measured within $ab$-plane without
 magnetic field (solid lines) and for NSMO film measured under the
 magnetic field of 50~kOe (dashed line). (b)Relation between
 magnetization and  $c/a$. The magnetization was measured at 5~K under 
 the magnetic field of 500~Oe after field cooling.}
\end{figure}
\begin{figure}
  \caption{\label{fig2}Temperature dependence of the out-of-plane
 resistivity of the NSMO/La-STO film under the magnetic field of 0 and
 50~kOe applied parallel to the plane. The inset shows the schematic
 view of the patterned structure for the measurement.} 
\end{figure}
\begin{figure}
  \caption{\label{fig3}Energy dependence of the normalized intensity of the
 interference term at azimuthal angle $\Psi=90^{\circ}$ for the NSMO
 film, measured at room temperature (triangles) and at 150~K (circles). 
 Solid line shows calculated result with the parameter $\Delta E=1.45$~eV
 ($\Delta E$ is the energy split of Mn $4p$ levels). The error bar at
 the bottom of the dip represents one standard deviation of the data.}
\end{figure}
\begin{figure}
 \caption{\label{fig4}Relation between $c/a$ and magnitude of Mn $4p$
 energy splitting. Solid line represents $\Delta E_{\mathrm{C}}/\Delta E_{\mathrm{JT}}=0$. Two arrows indicate the $\Delta E$ values used to obtain the worst possible value of $\Delta E_{\mathrm{C}}/\Delta E_{\mathrm{JT}}$. 
 The inset shows the azimuthal dependence of the interference term for
 the NSMO film measured at room temperature. The solid curve in the
 inset is calculated with  $\Delta E=1.45$~eV.}
\end{figure}

\begin{thebibliography}{99}
\bibitem{note}Other data points could be used in principle to further
	 reduce the upper limit of $\Delta E_{\mathrm{C}}/\Delta E_{\mathrm{JT}}$. However, they correspond to the films in the paramagnetic state for which the currently available calculation is less reliable.
\end{thebibliography}
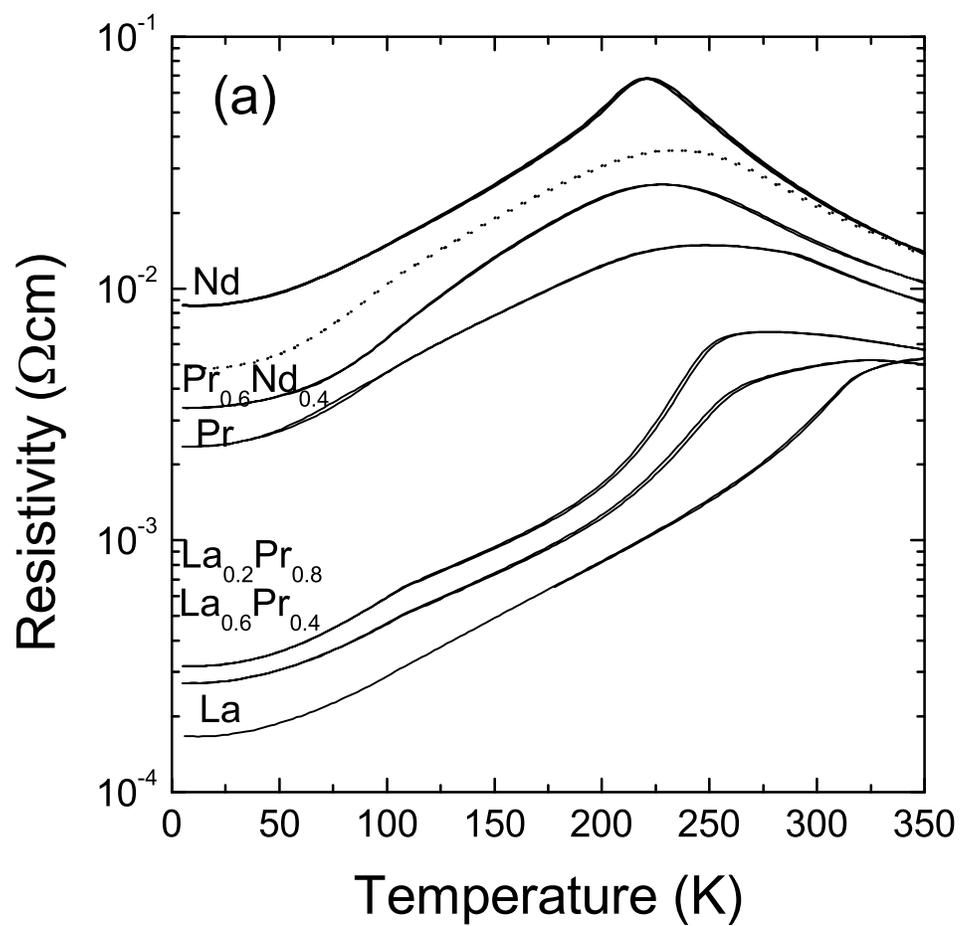
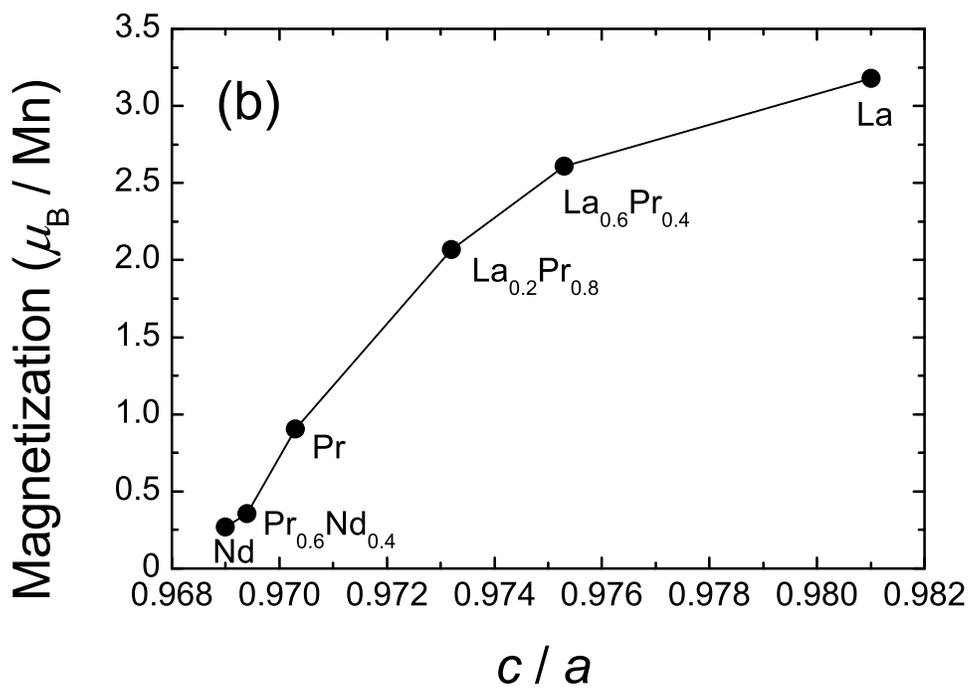
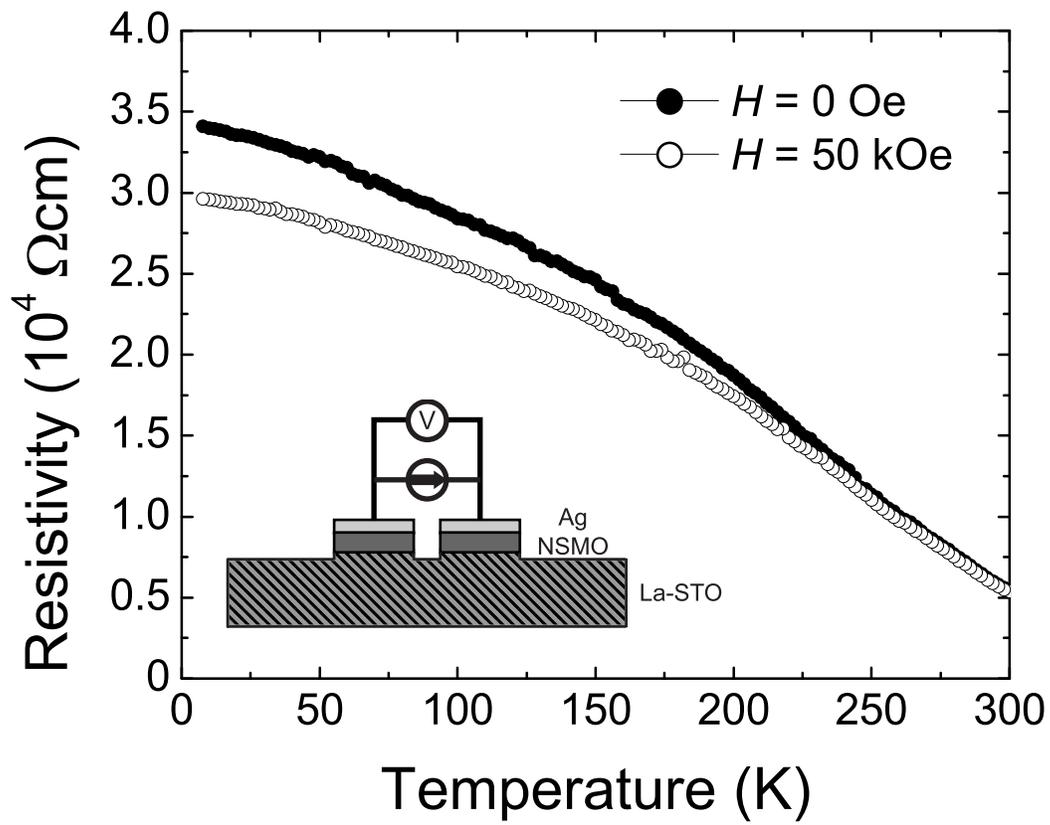
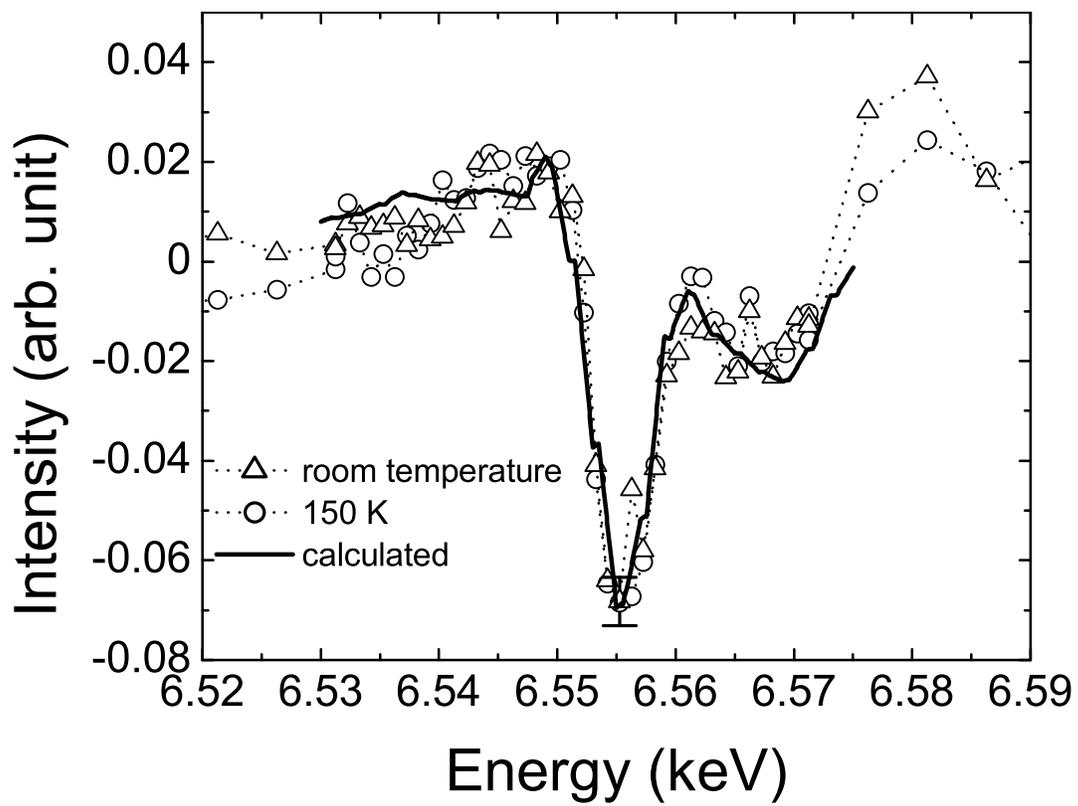
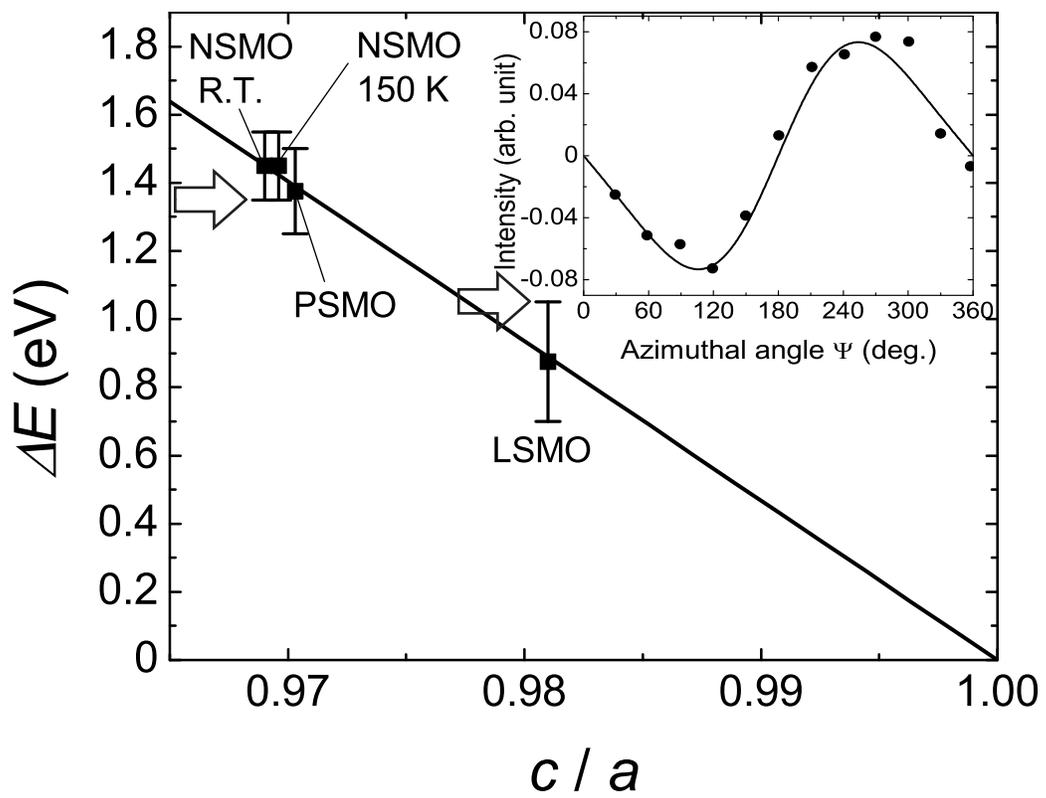
\end{document}